\documentclass[11pt]{article}

\usepackage{amsmath}

\newcommand{\N}{N\raise.7ex\hbox{\underline{$\circ $}}$\;$}
\begin{document}

\title{The Runge-Lenz vector for quantum Kepler problem in the space of  positive constant
curvature and complex parabolic coordinates
 }
\author{ A.A. Bogush, V.S. Otchik, V.M. Red'kov
\\
{\small    Institute of Physics,  National Academy of Sciences
of Belarus  }\\
{\small Institute for Command Engineers,     Ministry for
Emergency Services of Belarus } }
\date{}
\maketitle
\begin{abstract}

By analogy with the Lobachevsky space $H_{3}$,  generalized
parabolic coordinates  $(t_{1},t_{2},\phi)$ are introduced in
Riemannian space model of positive constant  curvature $S_{3}$. In
this case parabolic coordinates turn  out to be complex-valued and
obey additional restrictions involving  the complex conjugation.
In that  complex  coordinate system,  the  quantum-mechanical
Coulomb problem is  studied: separation of variables is  carried
out  and   the wave solutions in terms of hypergeometric functions
are obtained. At separating the variables, two parameters $k_{1}$
and $k_{2}$ are introduced, and an operator $B$ with the
 eigenvalues $(k_{1}+k_{2})$ is found, which is related to third
component  of the known Runge-Lenz vector in space $S_{3}$ as
follows: $ i  B = A _{3} + i \vec{L}^{2}$,  whereas  in the
Lobachevsky space as $B =   A _{3} +  \vec{L}^{2} $. General
aspects of  the possibility to employ complex coordinate systems
in the  real space model $S_{3}$ are discussed.

\end{abstract}

\section{Introduction}

In Euclidean 3-dimension space $E_{3}$ there exist 11  coordinate systems [1-3],
 allowing for the complete  separation of variables in the Helmholtz  equation
\begin{eqnarray}
[ {1 \over \sqrt{g}} \; {\partial \over \partial x^{\alpha}} \; \sqrt{g}
\; g^{\alpha \beta}\; {\partial \over \partial x^{\alpha}} \;+ \;
 \mbox{const}  \;  ]\;   \Phi (x^{1}, x^{2}, x^{3}) = 0 \;\; ;
\label{1.1'}
\end{eqnarray}

\noindent $g^{\alpha \beta}(x)$ stands for the metric tensor of space $E_{3}$ specified
for  curvilinear coordinates $(x^{1},x^{2},x^{3})$.
Solution of the same problem for
spaces of constant positive and negative curvature, Riemannian  $S_{3}$ and Lobachevsky $H_{3}$
 models  was given by Olevsky in [4], see also [5]. It was established that  there exist 34
 such  coordinate systems for hyperbolic space  $H_{3}$, whereas in the case of spherical model
 $S_{3}$  the number  of those systems is only  6.

 This result may seem rather unexpected and  even intriguing by some reasons.
Indeed,  there must exist a limiting procedure from  both curved space models $H_{3}$ and
$S_{3}$ to  flat space:
$$
H_{3} , \; S_{3} \qquad \longrightarrow \qquad E_{3}
$$

\noindent and it is natural  to expect the reduction $34
\longrightarrow 11$, however one  can hardly perform the reduction
$6 \longrightarrow 11 $.

The above asymmetry between  $H_{3}$ and $S_{3}$  may be seen   as even more   strange  if
one  calls the known relations   of these models  through  the analytical continuation:
\begin{eqnarray}
S_{3} \qquad  u_{0}^{2} +  u_{1}^{2}+ u_{2}^{2} + u_{3}^{2}  = + R^{2} \; ,
\nonumber
\\
H_{3} \qquad  u_{0}^{2} -  u_{1}^{2}- u_{2}^{2} - u_{3}^{2}  = + R^{2} \; ;
\label{a}
\end{eqnarray}

\noindent  where $R$ stands for a curvature radius for $S_{3}$ and $H_{3}$.

The  asymmetry  of the models $H_{3}$ and $S_{3}$  with respect to
coordinate systems finds its logical  corollary  when turning to the study of
the quantum mechanical  model for a hydrogen atom  on the background of  a curved space.
Firstly,
such a model was considered  in [6,7,8] where  the wave function in spherical coordinates
and energy spectrum were established. In particular,  an additional degeneracy
like in the  case of flat space was observed,
 which presumes  existence of  a hidden symmetry in the (curved space)  problem.
 In [9-12] the symmetry operators accounting for such additional degeneracy in Kepler
  problem on curved space background ware found for both model $H_{3}$ and $S_{3}$,
 and    an analog
  of the conventional Runge-Lenz vector in the flat space  was constructed.

  Connection between  the Runge-Lenz operator $\vec{A}$  in the quantum Kepler problem
and parabolic coordinates in   Euclidean space is well known: by
 solving the   Schr\"{o}dinger equation in these coordinates  the  eigenfunctions of
 the third component  $A_{3}$
   arise [13]. Analogous situation exists in the hyperbolic
    space $H_{3}$  but not in  in the spherical $S_{3}$ [14]. In the Lobachewsky space,
    among 34 coordinates established by Olevsky one may select  one  special case,
 parabolic     system of coordinates in $H_{3}$, in which  the Schr\"{o}dinger equation  allows the separation of variables and the  wave
 functions arisen turn out  to be  eigenfunctions of the operator $B = A_{3} +L^{2}$.
 Among six coordinate systems mentioned  in [4] an analog of parabolic coordinates
  is  not encountered.

If one looks at  34 and 6  systems in $H_{3}$ and $S_{3}$ respectively, one can note that
all
six ones from $S_{3}$ have their counterparts in $H_{3}$.
    The  main  purpose   of  the present paper consists is the search  of some counterparts
    of remaining $34-6=28$  systems. It turns  out that such 28  systems in $S_{3}$
     can be constructed, but they  should be complex-valued; to preserve real nature of the geometrical space  one must impose
    additional  restrictions  including complex conjugation.

In particular, the complex  analog for parabolic coordinates in space  of the positive
curvature $S_{3}$  can be introduced and used in studying the quantum mechanical
Kepler problem in this space.

\section{Complex parabolic coordinates in real space $S_{3}$}

Let us start with the following fact: from the the metric in Lobachevsky space
\begin{eqnarray}
dl^{2} = R^{2} \; [ \; d \chi ^{2} + \sinh ^{2} \chi \; (d
\theta ^{2} + \sin ^{2} \theta \; d\phi ^{2})\;]   \; , \;\; \chi
\in [0, + \infty )
\label{1.2a}
\end{eqnarray}

\noindent   by means of
the formal change
 $ \chi \; \rightarrow \; i \chi \; , \; \; \sinh \chi \; \longrightarrow
\; i \sin \chi \; $ one can  obtain the corresponding metric of
the Riemannian space \begin{eqnarray}
 dl^{2} = - R^{2} \; [ \; d
\chi ^{2} + \sin ^{2} \chi \; (d \theta ^{2} + \sin ^{2} \theta \;
d\phi ^{2})\;] \; , \;\; \;\; НО \;\;\chi \in [0, \pi ] \; .
\label{1.2b}
\end{eqnarray}

\noindent
This  simple observation on $H_{3}-S_{3}$ connection  leads us to
interesting    consequences. Indeed,  let us
compare, for instance,   wave functions and spectra   for
hydrogen atom in spaces of negative and constant curvature
[14]:
\begin{eqnarray}
\underline{H_{3}}, \qquad \qquad
 \Psi _{nlm} (\chi, \theta, \phi ) = N \;
S(\chi) \; Y_{lm}(\theta , \phi ) \; \; ,
\nonumber
\\
S(\chi) = \sinh ^{l} \chi \; \exp [(n-l-1 - {e \over n}) \chi ]\;  \times
\nonumber
\\
F({e \over n} + l + 1 , l - n + 1, 2l+2; 1 - e^{-2\chi}) \; ,
\label{1.3a}
\\
\epsilon_{n} = - {e^{2} \over 2 n^{2} } - {1 \over 2} \; (n^{2}
-1) \; ;
\label{1.3b}
\end{eqnarray}
\begin{eqnarray}
\underline{S_{3} }, \qquad \qquad
\Psi _{nlm} (\chi, \theta, \phi ) =
K \; S(\chi) \; Y_{lm}(\theta , \phi ) \; \; ,
\nonumber
\\
S(\chi) = \sin ^{l} \chi \; \exp [(i (n-l-1) - {e \over n}) \chi] \;  \times
\nonumber
\\
F(-i {e \over n} + l + 1 , l - n + 1, 2l+2;\; 1 - e^{-2i\chi}) \;
,
\label{1.4a}
\\
\epsilon_{n} = - {e^{2} \over 2 n^{2} } + {1 \over 2} \; (n^{2}
-1) \; \; , \;\; e = {\alpha \over  R } /  { M \hbar^{2} \over
R^{2} } \; ;
 \label{1.4b}
\end{eqnarray}

\noindent
quantity ($M\hbar^{2} /R^{2}$ provides us with natural unit for energy,
$e$ is a dimensionless parameter  characterizing intensity of the Coulomb interaction.
One may readily note that  these two solutions turn into each other
by the following formal substitutions:
\begin{eqnarray}
\chi \; \longrightarrow \; i \; \chi \; , \;\; e \longrightarrow
-i\; e \; , \;\; \epsilon \longrightarrow - \; \epsilon \; .
\label{1.5}
\end{eqnarray}

This example indicates that  the relation  between
$H_{3}$ and  $S_{3}$ reflected by substitution   $\chi \rightarrow i \chi$ is  meaningful.
In the context  of the described  above situation with coordinate systems in
$H_{3}$ and $S_{3}$,  let us  make use of this correspondence
($ \chi \; \longrightarrow \; i \; \chi$) as follows:

\begin{quotation}

 Let in the Lobachevsky space  $H_{3}$ be chosen  a coordinate
system $(\rho_{1},\; \rho_{2}, \; \rho_{3} )$
(one of those 34 found by Olevsky), then as a first step one has to  establish
connection of  such a system with spherical one:
\begin{eqnarray}
\rho _{k} = f_{k}( \chi, \; \theta, \; \phi ) \; ,
\label{1.6a}
\end{eqnarray}

\noindent and  a second step  is to  introduce a corresponding coordinate
system in the space $S_{3}$ through the formal change
$\chi \longrightarrow i\;\chi$:
\begin{eqnarray}
\rho _{k} = f_{k}( i \chi, \; \theta, \; \phi ) \; .
\label{1.6b}
\end{eqnarray}

With  help of this prescription one can determine 34 coordinate systems  in space $S_{3}$ in
comparison  to six ones given in  [4]. It turns out that  28 new (added) coordinate systems
are complex-valued and  therefore  additional restrictions  should be imposed
 which involve  complex conjugation.
All these  extra  coordinate systems  permit the full separation of
variables in the Helmholtz  equation on the sphere $S_{3}$.

\end{quotation}

Below, only one example of such coordinates, analog of   the parabolic ones  in space $H_{3}$,
will be examined in detail  and applied to the study of the quantum-mechanical Kepler problem on the
sphere $S_{3}$.

\section{Complex parabolic coordinates in $S_{3}$ and \\the hydrogen atom }

In  [4] Olevsky had given  the following coordinate system (the case  $XXV$)
in Lobachevsky space:
\begin{eqnarray}
dl^{2} =  \;R^{2} \left [ {(\rho_{1} - \rho_{2}) \over 4 (\rho _{1} -
a)(\rho_{1} - b)^{2}} \; d \rho^{2}_{1} \; + \; {(\rho_{2} -
\rho_{1}) \over 4 (\rho _{2} - a)(\rho_{2} - b)^{2}} \; d
\rho^{2}_{2} \; - \; (\rho_{1} - a)(\rho_{2} - a) d \rho_{3}^{2} \right ]
\;   ,
\label{2.1a}
\end{eqnarray}

\noindent    where  $(\rho_{1},\; \rho_{2}, \; \rho_{3} )$ are connected with
the four "Cartesian" \hspace{2mm}  (dimensionless) coordinates  $(x_{0}, x_{1},x_{2}, x_{3})$
by the formulas
\begin{eqnarray}
x_{0}^{2} - x_{1}^{2} - x_{2}^{2} - x_{3} ^{2} = 1 \;
,\;\;  x_{0}  > +1 \; ,
\nonumber
\\
{x_{2} \over x_{1}} = \tan [(a-b) \rho_{3}] \; , \;\;
b < \rho_{1} < a < \rho_{2} \; ,
\nonumber
\\
{x_{1}^{2} +  x_{2}^{2}  \over \rho_{i} - a } + {x_{3}^{2} -
x_{0}^{2}  \over \rho_{i} - b } + {(x_{3} - x_{0})^{2} \over (\rho
_{i}  - b )^{2}} = 0 \; , \; (i = 1,2 )\; .
\label{2.1b}
\end{eqnarray}

\noindent
With $a = +1, \; b = 0 \;$  and notations
$
x_{1}^{2} +  x_{2}^{2}  = \sigma^{2} \; , \;\;
x_{3} - x_{0} = U \; , \;\;
x_{3} + x_{0} = V \; ,
$
eqs. (\ref{2.1b}) give
\begin{eqnarray}
\sigma^{2} + UV = -1 \; , \;\; x_{1} = \sigma \cos \rho_{3} \; , \;\;
x_{2} = \sigma \sin \rho_{3} \; ,
\nonumber
\\
{ \sigma^{2} \over \rho_{1} -1 } + {UV \over \rho_{1}} +
{U^{2} \over \rho_{1}^{2}} = 0 \; , \;\;\;\;
{ \sigma^{2} \over \rho_{2} -1 } + {UV \over \rho_{2}} +
{U^{2} \over \rho_{2}^{2}} = 0 \; ,
\nonumber
\end{eqnarray}

\noindent  Combining two last relations one  obtains
\begin{eqnarray}
( {\rho_{1} \over \rho_{1} -1 }  -  {\rho_{2} \over \rho_{2} -1 } )
\sigma^{2}  + ( {1 \over \rho_{1} }  -  {1 \over \rho_{2} } )
 U^{2} = 0\; ,  \;
( {\rho_{1}^{2} \over \rho_{1} -1 }  - {\rho_{2}^{2} \over \rho_{2} -1 } )
\sigma^{2}  + ( \rho_{1}  - \rho_{2} ) U V = 0  \; ,
\nonumber
\end{eqnarray}

\noindent  whence having  in mind  $\sigma^{2} + UV = -1$, one gets
\begin{eqnarray}
{U \over V} = { \rho_{1} \rho_{2} \over \rho_{1} \rho_{2} -
\rho_{1} -  \rho_{2}  } \; , \;\; UV =
\rho_{1} \rho_{2} - \rho_{1} -  \rho_{2}  \; ,
\nonumber
\end{eqnarray}

\noindent
and further
\begin{eqnarray}
U^{2} = \rho_{1} \rho_{2} \; , \;\; V = U \; {
\rho_{1} \rho_{2} - \rho_{1} -  \rho_{2}  \over \rho_{1} \rho_{2} } \; .
\nonumber
\end{eqnarray}

\noindent As a result, for  $U,\; V, \; \sigma $ we  arrive at (take notice that
the Lobachevsky model
is realized  on the  upper sheet of hyperboloid $x_{0} > +1$,
and  therefore  $x_{3} - x_{0} \leq  0\; $)
\begin{eqnarray}
u = x_{3} - x_{0} = - \sqrt{\rho_{1} \rho_{2}} \; , \;\;
V = x_{3} + x_{0} = { \rho_{1} +  \rho_{2} - \rho_{1} \rho_{2} \over
\sqrt{ \rho_{1} \rho_{2} } }   \; ,
\nonumber
\\
\sigma = \sqrt{-1 - UV }= \sqrt{-(1-\rho_{1})(1- \rho_{2})} \; .
\nonumber
\end{eqnarray}

\noindent Thus, explicit formulas relating
$\rho_{1},\rho_{2},\rho_{3}$  with  Cartesian
coordinates $(_{0},x_{l}$ look as
\begin{eqnarray}
x_{1} =
\sqrt{-(1-\rho_{1})(1-\rho_{2})}\; \cos  \rho_{3} \;\; ,\;\; x_{2}
= \sqrt{-(1-\rho_{1})(1-\rho_{2})}\; \sin  \rho_{3} \;\; ,
\nonumber
\\
x_{3} = {\rho_{1} + \rho_{2} - 2 \rho_{1} \rho_{2} \over 2
\sqrt{\rho_{1} \rho_{2} }} \;\;  , \;\; x_{0} = {\rho_{1} +
\rho_{2} \over  2  \sqrt{ \rho_{1} \rho_{2} } }\; ;
\label{2.2}
\end{eqnarray}

\noindent and the  inverse formulas are
\begin{eqnarray}
\rho_{1} = {x_{0} - x_{3}  \over  x_{0} +  x } \;\; , \;\;
\rho_{2} = {x_{0} - x_{3}  \over  x_{0} -  x } \;\; , \;\;
\rho_{3} = \mbox{arctan}\; { x_{2}  \over  x_{1} } \;\;      , \;\; x =
\sqrt{x_{1}^{2} + x_{2}^{2} + x_{3}^{2}}\; .
\label{2.3}
\end{eqnarray}

Now, instead of the introduced $\rho_{1},\rho_{2},\rho_{3}$ one can   define
other coordinates which  behave    simply
 in the limit  $R \rightarrow \infty $  (the curvature vanishes). Such a limiting
  procedure    for spherical  coordinates of  the hyperbolic space $H_{3}$ with metric
\begin{eqnarray}
dl^{2} = \rho ^{2} \; [ \; d \chi ^{2} + \sinh ^{2} \chi \;
(d \theta ^{2} + \sin ^{2} \theta \; d\phi ^{2})\;]
\nonumber
\end{eqnarray}

\noindent  going over into spherical ones of the flat space $E_{3}$
\begin{eqnarray}
dl^{2} = [ \; d r  ^{2} + r^{2} \;
(d \theta ^{2} + \sin ^{2} \theta \; d \phi ^{2})\;] \; ,
\nonumber
\end{eqnarray}

\noindent looks as follows:
\begin{eqnarray}
\lim_{R \rightarrow \infty} (R \chi ) = r \;\; , \;\;
\lim_{R \rightarrow \infty} (R \sinh \chi ) = r \;\; .
\label{2.4a}
\end{eqnarray}

Eliminating   $x_{0}$ through  $q_{l}$:
\begin{eqnarray}
q_{l} = {x_{l} \over x_{0}} = {x_{l} \over + \sqrt{1 + x^{2}}} \;\; , \qquad
q_{l} = \tanh \chi \; n_{l} \; ,\nonumber
\\
 n_{l} = (\sin \theta \cos \phi,
\sin \theta \sin \phi, \cos \theta )\;, \; q =
\sqrt{q_{1}^{2} + q_{2}^{2} + q_{3}^{2}} = \tan \chi \; ;
\nonumber
\end{eqnarray}

\noindent we can readily see  that when $R \rightarrow \infty$ the coordinates
$q_{l}$ will reduce to
\begin{eqnarray}
\lim_{R \rightarrow \infty } (R q_{l}) = \lim_{R
\rightarrow \infty } (R \tanh \chi \; n_{l}) = r n_{l} \; .
\label{2.4b}
\end{eqnarray}

\noindent So, to have coordinates  with  the known and understandable behavior
in the limit $R \rightarrow \infty $ we  define   coordinates
$t_{1},t_{2},\phi$:
\begin{eqnarray}
t_{1} = 1 - \rho_{1} = {q_{3} + q \over 1 + q }\;\; , \;\; t_{2} =
1 - \rho_{2} = {q_{3} - q \over 1 - q } \;\; , \;\; \phi =
\rho_{3} = \mbox{arctan} \; {q_{2} \over q_{1}} \; ;
\label{2.5a}
\end{eqnarray}

\noindent in the limit of the flat space they provide us with  the known parabolic  coordinates
$(\xi, \eta, \phi)$ (see in [13]):
\begin{eqnarray}
\lim_{R \rightarrow \infty } (R t_{1}) = z + r = \xi \;\; ,
\;\; \lim_{R \rightarrow \infty } (R t_{2}) = z - r = - \eta
\;\; .
\label{2.5b}
\end{eqnarray}

\noindent
The metric  (\ref{2.1a})  in coordinates   $(t_{1},t_{2},\phi)$  takes the form
\begin{eqnarray}
dl^{2} =  R^{2}\; [ \; {t_{1} - t_{2} \over 4 t_{1} (1 - t_{1})^{2}} \;
dt_{1}^{2} \;+\;  {t_{2} - t_{1} \over 4 t_{2} (1 - t_{2})^{2}} \;
dt_{2}^{2}  \; - \; t_{1} t_{2} \; d \phi^{2} \; ] \; ,
\nonumber
\\
 0 \leq
t_{1} \leq 1 \, \, , \qquad   t_{2} \leq 0 \;\; , \qquad  0 \leq \phi
\leq 2\pi \; .
\label{2.6}
\end{eqnarray}

Now, with the help of the rules  (\ref{1.6a})-(\ref{1.6b}) one has to define
 corresponding parabolic coordinates $t_{1}, t_{2}$ on the sphere $S_{3}$.
To this end,  coordinates  $(t_{1}, t_{2})$ in  $H_{3}$ must be expressed in term of spherical ones
$(\chi, \theta)$:
\begin{eqnarray}
t_{1} = (1 + \cos \theta )\; { \tanh \chi \over 1 + \tanh \chi }
\; , \;\; t_{2} = (1 - \cos \theta )\; { - \tanh \chi \over 1 -
\tanh \chi } \; .
\label{2.7}
\end{eqnarray}

\noindent From  (\ref{2.7}) we get defining relations  for corresponding coordinates  in $S_{3}$
\begin{eqnarray}
t_{1} = (1 + \cos \theta )\; { i \tan \chi \over 1 + i \tan \chi }
\; , \;\; t_{2} = (1 - \cos \theta )\; { - i \tan \chi \over 1 - i
\tan \chi } \; .
\label{2.8a}
\end{eqnarray}

\noindent Take special notice that  $(t_{1}$ and $t_{2})$  in  (\ref{2.8a})
are complex-valued
expressed  through  two real   $(\chi, \theta)$.
The inverse formulas are readily found:
\begin{eqnarray}
1 + \cos \theta = t_{1} (1 + {1 \over i \tan \chi}) \;\; , \;\;
1 - \cos \theta = t_{1} (1 - {1 \over i \tan \chi }) \;\; , \;\;
\nonumber
\end{eqnarray}

\noindent and therefore
\begin{eqnarray}
\cos \theta  = {t_{1} - t_{2} - 2 t_{1} t_{2} \over    t_{1} -
t_{2} }\; \; , \;\;\;i \; \tan \chi  = {t_{1} - t_{2} \over 2 - t_{1} - t_{2} }
\; .
\label{2.8b}
\end{eqnarray}

\noindent
So defined  parametrization of  $S_{3}$ by  coordinates
$t_{1}, t_{2}$ can be additionally  detailed by the formulas:
\begin{eqnarray}
t_{1} =(1 + \cos \theta ) \; \varphi (\chi) \;\; , \;\;
t_{2} =(1 - \cos \theta ) \; \varphi^{*} (\chi) \;\; , \;\;
\nonumber
\\
\varphi (\chi) = \sin^{2} \chi + i \sin \chi \cos \chi = \sin \chi
\; \exp [ i({\pi \over 2} - \chi)] \; .
\label{2.9}
\end{eqnarray}

\noindent
From (\ref{2.9}) one can  derive the relationship between $t_{1}$ and $t_{2}$:
\begin{eqnarray}
t_{1}t_{1}^{*} = t_{1} -t_{2} - t_{1}t_{2} \;  .
\label{A}
\end{eqnarray}

\noindent its existence may  evidently  be referred  to the real nature of the space $S_{3}$.
The values of  $\theta = 0$ and $\theta = \pi$  are peculiar:
\begin{eqnarray}
\theta = 0 \; ,  \qquad  \Longrightarrow \qquad t_{1} = 2 \; \varphi (\chi) \; ,
\; t_{2} = 0 \;\; ;
\nonumber
\\
\theta = \pi  \; \qquad  \Longrightarrow \qquad  t_{1} = 0 \;\; ; \;\;
\;\; t_{2} = 2 \; \varphi ^{*} (\chi) \; .
\label{2.10}
\end{eqnarray}

\noindent
In the following, so defined coordinates
 $(t_{1},t_{2},\phi)$ are called parabolic  coordinates on the sphere $S_{3}$.
 In the limit of the flat space, they  reduce to  the ordinary parabolic coordinates [...]
 in accordance  with
 \begin{eqnarray}
 \lim_{\rho \rightarrow \infty } \; (-i \rho t_{1}) = \xi \;\; ,
\;\; \lim_{\rho \rightarrow \infty } \; (-i \rho t_{2}) = -\eta
\;\; .
\label{2.11}
\end{eqnarray}

Now, we transform  the metric of the space $S_{3}$ in spherical coordinates
\begin{eqnarray}
dl^{2} = - R ^{2} \; [ \; d \chi ^{2} + \sin ^{2} \chi \;
(d \theta ^{2} + \sin ^{2} \theta \; d\phi ^{2})\;]   \; , \;\;
\nonumber
\end{eqnarray}

\noindent to complex parabolic    $t_{1},t_{2},\phi$.
As a first step, with the help of
\begin{eqnarray}
\sin^{2} \theta =
t_{2} \; {1 - i q \over -iq } \;  t_{1} \; {1 + i q \over iq } \; = \;
t_{1} \; t_{2} \; {1 + q^{2} \over q^{2}} \;\; , \;\;\; \sin^{2} \chi =
{q^{2} \over 1 + q^{2} } \; ,
\nonumber
\end{eqnarray}

\noindent we get
\begin{eqnarray}
\sin^{2} \chi \; \sin^{2} \theta \; d \phi^{2} = t_{1}\; t_{2}\; d
\phi^{2} \; .
\nonumber
\end{eqnarray}

\noindent As a second step,   we  have
\begin{eqnarray}
( d \theta) ^{2} = { 1 \over \sin^{2} \theta } \; (d \; \cos \theta )^{2} =
{1 \over t_{1} t_{2} } \; {q^{2} \over 1 + q^{2}} \;
[ d  (
{t_{1} + t_{2} - 2 t_{1}t_{2} \over t_{1} - t_{2} }  )  ]^{2} \; ,
\nonumber
\end{eqnarray}

\noindent   or
\begin{eqnarray}
\sin^{2} \chi \; (d \theta)^{2} =
{q^{4} \over (1+q^{2})^{2}}\; {4 \over t_{1} t_{2} (t_{1} - t_{2})^{4}}\;
[\; t_{2}(t_{2} - 1) \; dt_{1} \;  -   \; t_{1}(t_{1} - 1) \; dt_{2}\; ]^{2} \; ;
\nonumber
\end{eqnarray}

\noindent so that
\begin{eqnarray}
\sin^{2} \chi \; (d \theta)^{2} = {1 \over 4(1- t_{1})^{2} (1-
t_{2})^{2} t_{1} t_{2} }\; [\; t_{2}\; (t_{2} - 1) \; dt_{1} \; -
\;  t_{1} \; (t_{1} - 1) \; dt_{2}\; ]^{2} \; .
\nonumber
\end{eqnarray}

\noindent Finally,  taking into account relations
\begin{eqnarray}
i \tan \chi = { t_{1} - t_{2} \over 2 - t_{1} - t_{2}} \; , \Longrightarrow \;
{i d \chi \over \cos^{2} \chi } =
[ \; dt_{1} \; {2(1 -t_{2}) \over  (2 - t_{1} - t_{2})^{2} } \; - \;
dt_{2} \; {2(1 -t_{1}) \over  ( 2 - t_{1} - t_{2})^{2} } \; ] \; \; ,
\nonumber
\end{eqnarray}

\noindent and
\begin{eqnarray}
\cos^{2} \chi = { 1 \over 1 + \tan^{2} \chi } = { (2 - t_{1} - t_{2})^{2} \over
4(1- t_{1}) (1- t_{2})} \; ,
\nonumber
\end{eqnarray}

\noindent we get
\begin{eqnarray}
(d \chi)^{2} = { -1 \over  4(1- t_{1})^{2} (1- t_{2})^{2} }\; [\;
(1-t_{2}) dt_{1} - (1 -t_{1}) dt_{2}\; ]^{2}  \; .
\nonumber
\end{eqnarray}

\noindent
Therefore, for the metric in parabolic  coordinates in $S_{3}$
 we have arrived at  the form
 \begin{eqnarray}
dl^{2} = -R^{2} \left [  \; {t_{2} - t_{1} \over 4 t_{1} (1 - t_{1})^{2}} \;
dt_{1}^{2} \;+\;  {t_{1} - t_{2} \over 4 t_{2} (1 - t_{2})^{2}} \;
dt_{2}^{2}  \; + \; t_{1} t_{2} d \phi^{2} \;\right ]  \; .
\label{2.13}
\end{eqnarray}

\noindent Formally, this formula differs from  its counterpart in the space $H_{3}$ only
by presence of $(-1)$ in  the expression for $dl^{2}$.

The Schr\"{o}dinger Hamiltonian for the Kepler problem
\begin{eqnarray}
H =  - {1 \over 2} {1 \over \sqrt{g}} \;{\partial \over \partial
x^{\alpha}}\; \sqrt{g}\; g^{\alpha \beta}\;{\partial \over
\partial x^{\beta}}\; \; - \; {e \over q} \;\; ,
\label{2.14a}
\end{eqnarray}

\noindent
will take the following explicit form
\begin{eqnarray}
H =  2 \; {1 - t_{1} \over t_{1} - t_{2} } \;
{\partial \over \partial t_{1}} t_{1}(1 - t_{1})
{\partial \over \partial t_{1}}   \; +
\;  2 \; {1 - t_{2} \over t_{2} - t_{1} } \;
{\partial \over \partial t_{2}} t_{2}(1 - t_{2})
{\partial \over \partial t_{2}}  \; - \;
\nonumber
\\
 - \;  {1 \over 2 t_{1} t_{2} }\;
{\partial^{2} \over \partial \phi^{2}} \; - \;
 i e \; {2 - t_{1} -t_{2}  \over  t_{1} -t_{2} } \;  \; .
\label{2.14c}
\end{eqnarray}

\noindent This Hamiltonian  may be referred to analogous one in $H_{3}$
with the help of formal changes:
$e \; \rightarrow \; -ie$ and    $H \; \rightarrow \; -H$.

Now, acting in the way used for  Lobachevsky space [14],
one can  perform the separation of variables in the Schr\"{o}dinger
equation in the spherical space $S_{3}$:
\begin{eqnarray}
\Psi (t_{1}, t_{2},\phi) = f_{1}(t_{1})\;
f_{2}(t_{2})\;e^{im\phi} \; .
\label{2.15}
\end{eqnarray}

\noindent From the equation  $H \Psi = \epsilon \; \Psi $  it follows
\begin{eqnarray}
f_{2} \; {2(1 - t_{1}) \over t_{1} - t_{2} } \; {d \over dt_{1}}\; t_{1}(1 -t_{1})\;
{d \over dt_{1}}\; f_{1} \; +
f_{1} \; {2(1 - t_{2}) \over t_{2} - t_{1} } \; {d \over dt_{2}}\; t_{2}(1 -t_{2})\;
{d \over dt_{2}}\; f_{2} \;
\nonumber
\\
+ \; { m^{2} \over 2 t_{1} t_{2}}\; f_{1} f_{2} \; - \; ie \; {2 -
t_{1} -t_{2}  \over  t_{1} -t_{2} } \; f_{1} f_{2}  = \epsilon \;
f_{1} f_{2} \; .
\label{2.16}
\end{eqnarray}

\noindent or
\begin{eqnarray}
{1 \over f_{1}} \;(1 -t_{1}) \; {d \over dt_{1}} \; t_{1}(1 -t_{1})\;{d\over dt_{1}}\;
f_{1} \; - {m^{2} \over 4 t_{1}} \; + {ie\over 2} \; t_{1} \; -
\; {\epsilon \over 2}\; t_{1} \;
\nonumber
\\
{1 \over f_{2}} \;(1 -t_{2}) \; {d \over dt_{2}} \; t_{2}(1
-t_{2})\;{d\over dt_{2}}\; f_{2} \; + {m^{2} \over 4 t_{2}} \; +
{ie \over 2} \; t_{2} \; + \; {\epsilon \over 2}\; t_{2} \; + \;
(k_{1} - k_{2}) = 0 \; ;
\label{2.17a}
\end{eqnarray}

\noindent where two separation constants $k_{1}$ and
$k_{2}$ are introduced:
\begin{eqnarray}
 k_{1} - k_{2}  = -i\; e \;    .
\label{2.17b}
\end{eqnarray}

\noindent As a result, we arrive at the system of two 2-order ordinary differential equations
\begin{eqnarray}
(1 -t_{1}) \; {d \over dt_{1}} \; t_{1}(1 -t_{1})\;{d\over dt_{1}}\;
f_{1} \;  +  \;   (
\; {ie - \epsilon \over 2}\; t_{1} \; - \; {m^{2} \over 4 t_{1}} \; + \; k_{1}
\;) f_{1} = 0 \;\; ,
\nonumber
\\
(1 -t_{2}) \; {d \over dt_{2}} \; t_{2}(1 -t_{2})\;{d\over
dt_{2}}\; f_{2} \;   + \;  ( \; {- ie - \epsilon \over 2}\; t_{2}
\; - \; {m^{2} \over 4 t_{1}} \; + \; k_{2} \; ) f_{2} = 0 \;\; .
\label{2.18}
\end{eqnarray}

\noindent Analogous system of equation in the Lobachevsky space
has the form [14]
\begin{eqnarray}
(1 - t_{1}) \; {d \over dt_{1}} \; t_{1}(1 - t_{1})\;{d \over dt_{1}}\;
f_{1} \;  +  \;   (
\; {-e + \epsilon \over 2}\; t_{1} \; - \; {m^{2} \over 4 t_{1}} \; + \; k_{1}
\; ) f_{1} = 0 \;\; ,
\nonumber
\\
(1 - t_{2}) \; {d \over dt_{2}} \; t_{2}(1 - t_{2})\;{d\over
dt_{2}}\; f_{2} \;   + \;   ( \; { e + \epsilon \over 2}\; t_{2}
\; - \; {m^{2} \over 4 t_{1}} \; + \; k_{2} \; ) f_{2} = 0 \;\; .
\label{2.19}
\end{eqnarray}

\noindent
Solutions of eqs.  (\ref{2.18}) and  (\ref{2.19}) can be searched   for
with the help of substitution
\begin{eqnarray}
f_{1} = t_{1}^{a_{1}} \; (1 - t_{1})^{b_{1}} \; S_{1}(t_{1}) \;\;
, \;\;\; f_{2} = t_{2}^{a_{2}} \; (1 - t_{2})^{b_{2}} \;
S_{2}(t_{2}) \;\; .
\label{2.20}
\end{eqnarray}

\noindent Below,  all calculations will be  done for the case
(\ref{2.18}); at any point, transition to $H_{3}$ space is accomplished by the formal
changes $\epsilon \; \rightarrow \; -\epsilon,\;\; -ie  \; \rightarrow \; e$.
It suffices to consider in detail only  the first equation for $f_{1}(t_{1})$
(index 1 is omitted below)
\begin{eqnarray}
t (1 - t) \; S'' + S' \; [2a (1 -t)  - 2bt + (1 - 2t) ] \;
\nonumber
\\
+ \; [\;  a(a-1) ({1\over t} - 1) - 2ab + b(b-1)({1\over t} - 1) +
a({1\over t} - 2) - b (2 - {1 \over 1 - t})
\nonumber
\\
+\;  {ie - \epsilon \over 2}({1 \over 1 -t } -1) - {m^{2} \over 4}
({1 \over t} + {1 \over 1 - t}) + k{1 \over 1-t}\; ]\; S = 0 \; .
\label{2.21}
\end{eqnarray}

\noindent   Both terms proportional to $t^{-1}$ and $(1 - t)^{-1}$ may be
 eliminated from the equation by adding the requirements:
\begin{eqnarray}
 a^{2} - {m^{2} \over 4} = 0 \;\; , \;\;\; b^{2} + {ie - \epsilon
\over 2} -{m^{2} \over 4} + k = 0 \; ;
\label{2.22}
\end{eqnarray}

\noindent then eq.  (\ref{2.21}) results in
\begin{eqnarray}
t (1 - t) \; S'' + S' \; [\; (2a + 1)  - (2a +  2b + 2 ) t\; ]
\;
\nonumber
\\
- \;[ a(a+1) + 2ab +  b(b+1) +  {ie - \epsilon \over 2} ]\; S = 0
\; .
\label{2.23}
\end{eqnarray}

\noindent
That is,   $S(t)$ turns out to be  a hypergeometric  function
 $S(t) = F(\alpha, \; \beta, \; \gamma ; \; t)$ whose parameters  are determined by
\begin{eqnarray}
\alpha + \beta + 1 = 2a + 2b + 2 \;\; , \qquad
\nonumber
\\
\alpha \beta =
a(a+1) + 2ab +  b(b+1) +  {ie - \epsilon \over 2} \; , \;\; \gamma
= 2a+1 \; ,
\label{2-2}
\end{eqnarray}

\noindent  which implies
\begin{eqnarray}\alpha = a + b  {1\over 2} + \sqrt{{1\over 4} + {\epsilon -ie
\over 2}} \; , \;\; \beta = a + b  {1\over 2} - \sqrt{{1\over 4} +
{\epsilon -ie \over 2}} \; , \;\; \gamma = 2a + 1 \; .
\label{2-3}
\end{eqnarray}

Thus,  the quantum mechanical Kepler problem in complex parabolic
 coordinates  $t_{1}, t_{2},\phi)$ has been solved in hypergeometric functions.
 The separation  constants $k_{1}$ and  $k_{2}$  are connected by the relation (\ref{2.17b}):
 \begin{eqnarray}
f_{1} = t_{1}^{a_{1}}\; (1 - t_{1})^{b_{1}}\;S_{1} \; , \;\;
f_{2} = t_{2}^{a_{2} }\; (1 - t_{2})^{b_{2}}\;S_{2} \;  ;
\nonumber
\\
S_{1} = F(\alpha_{1}, \; \beta_{1},\;\gamma_{1}; \; t_{1}) \; , \;\;
S_{2} = F(\alpha_{2}, \; \beta_{2},\;\gamma_{2}; \; t_{2}) \; ;
\nonumber
\\
a_{1}^{2} = {m^{2} \over 4 }  \; , \;\;
a_{2}^{2} = {m^{2} \over 4 }  \; ;
\nonumber
\\
b_{1}^{2} = {\epsilon - i e \over 2} + {m^{2} \over 4} - k_{1}  \; , \;\;
b_{2}^{2} = {\epsilon + i e \over 2} + {m^{2} \over 4} - k_{2}  \; ;
\nonumber
\\
\alpha_{1} = a_{1} + b_{1} +  {1\over 2} + \sqrt{{1\over 4} +
{\epsilon -ie \over 2}} \; , \;\;
\alpha_{2} = a_{2} + b_{2} +  {1\over 2} + \sqrt{{1\over 4} +
{\epsilon + i e \over 2}}  \, \;
\nonumber
\\
\beta_{1} = a_{1} + b_{1} +  {1\over 2} - \sqrt{{1\over 4} +
{\epsilon -ie \over 2}}  \; , \;\;
\beta_{2} = a_{2} + b_{2} +  {1\over 2} - \sqrt{{1\over 4} +
{\epsilon + i e \over 2}}  \; ;
\nonumber
\\
\gamma_{1} = 2a_{1} + 1  \; ;  \;\; \gamma_{2} = 2a_{2} + 1 \; .
\label{2.25}
\end{eqnarray}

\noindent
Some additional study  is required to obtain the physical  wave solutions
of the Schr\"{o}dinger equation.

\section{
The Runge-Lenz vector in  $S_{3}$ and complex parabolic \\coordinates
}

At separating the variables in Schr\"{o}dinger equation two constants were introduced
$k_{1}$ и $k_{2}$; the problem is to find an operator
that is diagonalized on wave functions
(\ref{2.15}) with eigenvalues
 $(k_{1}+k_{2})$:
\begin{eqnarray}
\hat{B} \; f_{1} \; f_{2} \; e^{im\phi} = ( k_{1}+k_{2} ) \;
f_{1} \; f_{2} \; e^{im\phi}  \; .
\label{3.1}
\end{eqnarray}

\noindent Taking into account  (\ref{2.19}),
one can obtain  for the operator$\hat{B}$ the following representation:
\begin{eqnarray}
\hat{B} =   -(1 - t_{1}) \; {\partial \over \partial t_{1}} \; t_{1}
(1 - t_{1}) \; {\partial \over \partial t_{1}} \;   - t_{1} \;
{( - H + ie ) \over 2}  - {1 \over 4 t_{1}} \; {\partial^{2} \over \partial
\phi^{2}}
\nonumber
\\
\; -(1 - t_{2}) \;  {\partial \over \partial t_{2}} \; t_{2} (1
- t_{2})  \;  {\partial \over \partial t_{2}} \;   - t_{2} \; {( -
H - ie ) \over 2}    - {1 \over 4 t_{2}} \; {\partial^{2} \over
\partial \phi^{2}} \;  ,
\label{3.2a}
\end{eqnarray}

\noindent or after substituting the expression  for  $H$  from (\ref{2.14c})
\begin{eqnarray}
\hat{B} =
-ie \; {t_{1} + t_{2}  - 2t_{1} t_{2} \over t_{1} -t_{2}} \; +  \;
{2t_{2}(1-t_{1})(1 - 2 t_{1}) \over t_{1} - t_{2}} \;
{\partial \over \partial t_{1}} \; +  \;
{2t_{1}(1-t_{2})(1 - 2 t_{2}) \over t_{2} - t_{1}} \;
 {\partial \over \partial t_{2}} \; +
\nonumber
\\
+ \; {2 t_{1} t_{2} (1 - t_{1})^{2} \over t_{1} - t_{2}} \;
{\partial^{2} \over  \partial t_{1}^{2}} \; + \; {2 t_{2} t_{1} (1
- t_{2})^{2} \over t_{2} - t_{1}}  \; {\partial^{2} \over
\partial t_{2}^{2}} \; - \; {t_{1} + t_{2}  \over  2 t_{1} t_{2}}
\; {\partial^{2} \over \partial \phi^{2} } \;  ;
\label{3.2b}
\end{eqnarray}

\noindent note the identity
\begin{eqnarray}
-ie \; {t_{1} + t_{2}  - 2t_{1} t_{2} \over t_{1} -t_{2}}  = -ie\;
\cos \theta = -ie\; {q_{3} \over q } \; .
\label{3.3}
\end{eqnarray}

In  $H_{3}$ and $S_{3}$ the quantum mechanical Runge-Lenz operator
is constructed in term of momentum and orbital momentum
by the formula [11,12.]
\begin{eqnarray}
\vec{A} = e{\vec{q} \over q} + {1\over 2} \; ([\vec{L} \;\vec{P}] -
[\vec{P} \;\vec{L}]) \; ,
\label{3.5}
\end{eqnarray}

\noindent where
\begin{eqnarray}
\vec{P} = (P_{i}) , \; P_{i} =  -i (\delta_{ij}  \mp q_{i} q_{j}) \; {\partial \over
 \partial q_{j}} \;\; , \;\ \vec{L} = [ \vec{q} \;\vec{P}] \;\; ,
\label{3.6}
\end{eqnarray}

\noindent  upper sign corresponds to the model  $H_{3}$, lower corresponds to
 $S_{3}$ model; operators and $\vec{L}$ и $\vec{P}$  are measured in units
 $\hbar$ and  $\hbar / \rho$ respectively.
In correspondence with symmetry of  space models, the components of
 $\vec{P}, \; \vec{L}$ obey the commutation relations of
 Lie algebras  $so(3.1)$ and $so(4)$:
\begin{eqnarray}
[L_{a}, \; L_{b}] =i\; \epsilon _{abc} \; L_{c}\;\; , \;\; [L_{a},
\; P_{b}] =i \; \epsilon _{abc} \; P_{c}\;\; , \;\; [P_{a}, \;
P_{b}] =\pm i\; \epsilon _{abc}\; L_{c}\;\; .
\nonumber
\end{eqnarray}

\noindent
As in  the above expression for $\hat{B}$, specific term
$\;-ie\;q_{3} / q \; $  is presented,  (in the model  $H_{3}$ we see the term
 $\; e\;q_{3} / q \;$, it is natural to look for certain relationship   between
 $\hat{B}$ and  $A_{3}$ -- they look as follows (all details are omitted here)
\begin{eqnarray}
\mbox{in} \;\;\; H_{3} \; , \qquad \qquad \hat{B} =  (\; A \; + \;
\vec{L}^{2}\; ) \; ,
\nonumber
\\
\mbox{in} \;\;\; S_{3} \; , \qquad       \qquad i \; \hat{B} = (\; A \; +
\;i\; \vec{L}^{2}\; ) \; .
\label{3.25b}
\end{eqnarray}

\end{document}